\documentclass{aa}
\usepackage{graphicx}
\usepackage[varg]{txfonts}
\usepackage{cases}
\usepackage{natbib}
\usepackage[usenames,dvipsnames,svgnames,table]{xcolor}

\newcommand{\red}[1]{{\color{red}{#1}}}
\newcommand{\cyan}[1]{{\color{cyan}{#1}}}
\newcommand{\magen}[1]{{\color{magenta}{#1}}}

\begin{document}
\title{Moving structures in ultraviolet bright points: observations from Solar Orbiter/EUI}

\author{Dong~Li}

\institute{Key Laboratory of Dark Matter and Space Astronomy, Purple Mountain Observatory, CAS, Nanjing 210023, PR China \\
           \email{lidong@pmo.ac.cn} \\}
\date{Received; accepted}

\titlerunning{Moving structures in ultraviolet bright points}
\authorrunning{Dong~Li.}

\abstract {Moving structures have been detected in coronal bright
points and in a solar flare in active regions, which were
bi-directional, symmetrical, simultaneous, and quasi-periodic
\citep{Ning14,Ning16,Li16}. They could be regarded as observational
evidence of plasma outflows via magnetic reconnection.} {We explored
pairs of moving structures in fifteen ultraviolet bright points
(UBPs), which were observed in the quiet Sun or inside a small
active region on 19 November 2020.} {The UBPs were measured by the
High Resolution (HRI) Telescopes of the Extreme Ultraviolet Imager
(EUI) on board the Solar Orbiter (SolO) in two passbands, HRI$_{\rm
EUV}$~174~{\AA} and HRI$_{\rm Ly\alpha}$~1216~{\AA}. The pairs of
moving structures are identified in time-distance slices along
curved slits of UBPs, and their quasi-periods are determined from
the fast Fourier transform and wavelet analysis method.} {Moving
structures observed in ten UBPs as starting from their bright cores
and propagating toward two ends, are interpreted as diverging
motions of bi-directional moving structures. These moving structures
are also characterized by simultaneity and symmetry and in the case
of seven UBPs they exhibit quasi-periodicity. Moving structures seen
in another five UBPs as originating from double ends and moving
closer, and merging together, are manifested as converging motions.
A sympathetic UBP induced by the primary UBP is observed at the edge
of a small active region, and their moving structures also show the
converging motion.} {The diverging motions of bi-directional moving
structures could be generated by outflows after magnetic
reconnections. The converging motions of two moving structures might
be caused by inflows through the magnetic reconnection, or might be
interpreted as upflows driven by the chromospheric evaporation.}


\maketitle

\section{Introduction}
Coronal bright points (CBP) are ubiquitous small-scale enhanced
emissions in the lower corona over the Sun, which are a fundamental
class of solar activities \citep[see][for a recent
review]{Madjarska19}. They are easily observed in the
ultraviolet/extreme-ultraviolet (UV/EUV) and soft X-ray (SXR)
wavelengths \citep[e.g.,][]{Golub74,Li13,Zhang14,Alipour15,Mou18},
and often reveal loop-shaped features that connect magnetic fields
with opposite polarities in quiet-Sun or active regions, as well as
in coronal holes
\citep{Harvey93,Li12,Li13,Zhang12,Mou16,Galsgaard19}. They have a
length range of about 5$-$30~Mm
\citep{Golub74,Habbal90,Ueda10,Mou18}. Their duration covers a wide
range, varying from a few minutes to several hours or even longer
\citep{Golub76,Habbal81,Harvey93,Zhang01}. Their length sizes and
durations are often related to the evolution of magnetic fields in
the underlying photosphere \citep{Golub77,Yokoyama95,Alexander11}.
Quasi-periodic oscillations with periods from  minutes to hours are
also reported in CBPs \citep{Tian08,Samanta15}. CBPs can be found
everywhere at the solar atmosphere and appear significantly to
coincide with magnetic neutral lines, which should play an important
role for understanding the coronal heating
\citep{Priest94,Shibata07,Tian07,Hosseini21}.

Magnetic reconnection is a basic dynamical process of the energy
release in solar eruptions, i.e., solar flares
\citep{Priest02,Su13}, CBPs \citep{Priest94,Ning20}, transition
region explosive events \citep{Innes97,Innes15,Huang19}, plasma jets
\citep{Jiang13,Tian14,Tiwari19,Shen21}, blinkers
\citep{Chae00,Subramanian08}, and UV/EUV bursts
\citep{Peter14,Young18,Tian21}, et cetera. Thus, the Sun could
provide a broad range of sizes for the observational features of
magnetic reconnection, such as: reconnection inflows and outflows of
solar flares \citep{Hara11,Liu13} and CBPs
\citep{Yokoyama95,Ning14,Li16,Shokri22} in solar active regions,
bi-directional plasma jets of transition region explosive events
\citep{Innes97,Huang14,Li18}, the magnetic null point in CBPs
\citep{Sun12,Zhang12}, and the flux rope in solar active regions
\citep{Cheng13,Lip16} or in coronal mass ejections
\citep{Lin05,Cheng20}, et cetera. Those observational evidences are
seen in multiple layers of solar atmosphere, such as the corona, the
transition region, and the chromosphere, and benefit from imaging
and spectroscopic observations with high-spatial or high-cadence
resolutions
\citep[e.g.,][]{Peter14,Tian14,Tiwari19,Yan20,Chitta21a}.

In the Solar Orbiter \citep[SolO;][]{Muller20} era, using High
Resolutions (HRI) Telescopes of the Extreme Ultraviolet Imager
\citep[EUI;][]{Rochus20}, three groups of small-scale brightening
events in the quiet Sun are reported. One group is named as
`campfire' \citep{Berghmans21,Mandal21,Zhukov21}, it could be driven
by the component reconnection, e.g., loops reconnecting at coronal
heights \citep{Chen21,Panesar21}. The second one is `microjet',
which is mostly likely produced by the low height reconnection with
small-scale chromospheric fields. Moreover, its energy is roughly
equal to the energy released from the predicted nanoflare theory
\citep{Hou21}. The third one is called as `fast repeating jet', it
shows bi-directions or uni-direction at a very short time scale of
about 20~s \citep{Chitta21}. Those three groups of brightening
events are all associated with the reconnection model and reveal the
similar transient nature. They are regarded as the observational
signatures of small-scale magnetic reconnection in the quiet-Sun
region.

The classical 2-D reconnection model \citep{Sturrock64} suggested
that bi-directional outflows via magnetic reconnection could be
produced simultaneously, and their emissions along propagation paths
may be seen in the EUV and optical wavelengths, while the
reconnection flows were regarded as inflows or outflows. Based on
this idea, \cite{Ning14} investigated several pairs of moving
structures originating from the brightness core of a CBP, and they
were found to be characterized by bi-directions, symmetry,
simultaneity, and quasi-periodicity. \cite{Li16} further found that
a small pair of positive and negative magnetic fields moved
gradually closer to each other during bi-directional moving
structures in another CBP. Therefore, those bi-directional moving
structures were attributed to observational outflows after magnetic
reconnections. Similar bi-directional moving structures were also
found in a C-class flare \citep{Ning16} and in a small-scale loop
system \citep{Ning20}, confirming that they were results of the
small-scale magnetic reconnection. \cite{Zhang13} studied moving
structures between a primary CBP and its sympathetic CBPs. The
moving structures were found to be propagate upward along large
loop-like paths, and they were regarded as the signature of
chromosphere evaporation after magnetic reconnection. It should be
pointed out that all above CBPs/flares were observed inside active
regions on the Sun.

CBPs are a very specific class of events characterized by hot plasma
emissions and by a typical topology of the photospheric magnetic
field, such as the small-scale and new-emergence magnetic field that
often has opposite polarities
\citep{Golub76,Priest94,Alexander11,Li12}. In this study, the novel
SolO/EUI data obtained during cruise phase is used to investigate
bi-directional moving structures in quiet and active regions. We
only have images at EUI~174~{\AA} and 1216~{\AA}, but without images
at higher temperatures or photospheric magnetic field maps.
Therefore, the studied features cannot be identified as CBPs but are
simply named as UV bright points (UBPs). They most likely also sit
at the larger sizes, longer lifetimes end of the campfires'
distribution \citep[e.g.,][]{Berghmans21}. This article is organized
as follows: Section~2 describes the observation, and Section~3
presents our main results, Section~4 summaries the conclusion and
gives some discussions.

\section{Observation}
The observational data was captured by two HRI telescopes of
SolO/EUI, and the observed passbands are centered at 174~{\AA} and
1216~{\AA}, respectively. The first telescope named `HRI$_{\rm
EUV}$' is dominated by two emission lines of Fe IX and Fe X in the
lower corona, which has a formation temperature of about 1~MK. The
second one labeled `HRI$_{\rm Ly\alpha}$' is dominated by the
Ly$\alpha$ emission of neutral hydrogen line in the upper
chromosphere \citep{Berghmans21,Chen21}. The data studied here were
obtained on 19 November 2020 over about 3 hours. They are divided
into three sets, i.e., from 12:00~UT to 12:39~UT (\cyan{$\circ$}),
from 13:00~UT to 13:39~UT (\red{$\star$}) and from 14:00~UT to
14:39~UT (\magen{$\diamond$}), and each acquired at a time cadence
of 15~s. At the time of observations, SolO was located at
121.1$^{\circ}$ west in solar longitude from the Earth-Sun line.
Thus, it looked at a different side of the Sun, which it is not
observable with telescopes in Earth orbit, i.e., the Solar Dynamics
Observatory (SDO). The distance between SolO and the Sun was about
0.92~AU, thus the pixel scale was about 328~km~pixel$^{-1}$ for
HRI$_{\rm EUV}$ images, and it was about 686~km~pixel$^{-1}$ for
HRI$_{\rm Ly\alpha}$ images as these were binned 2$\times$2 aboard.

\begin{figure}[ht]
\centering
\includegraphics[width=\linewidth,clip=]{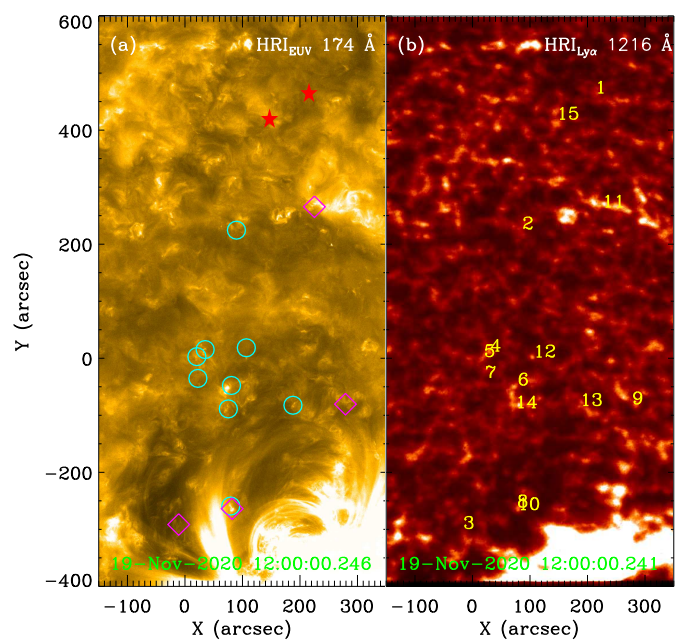}
\caption{Snapshots in HRI$_{\rm EUV}$ (a) and HRI$_{\rm Ly\alpha}$
(b) passbands measured by SolO/EUI. Fifteen UBPs are marked by
numbers, while cyan circles (\cyan{$\circ$}), red stars
(\red{$\star$}) and magenta diamonds (\magen{$\diamond$}) outline
their locations in three data sets, respectively. \label{snap}}
\end{figure}

\begin{figure*}[ht]
\centering
\includegraphics[width=0.93\linewidth,clip=]{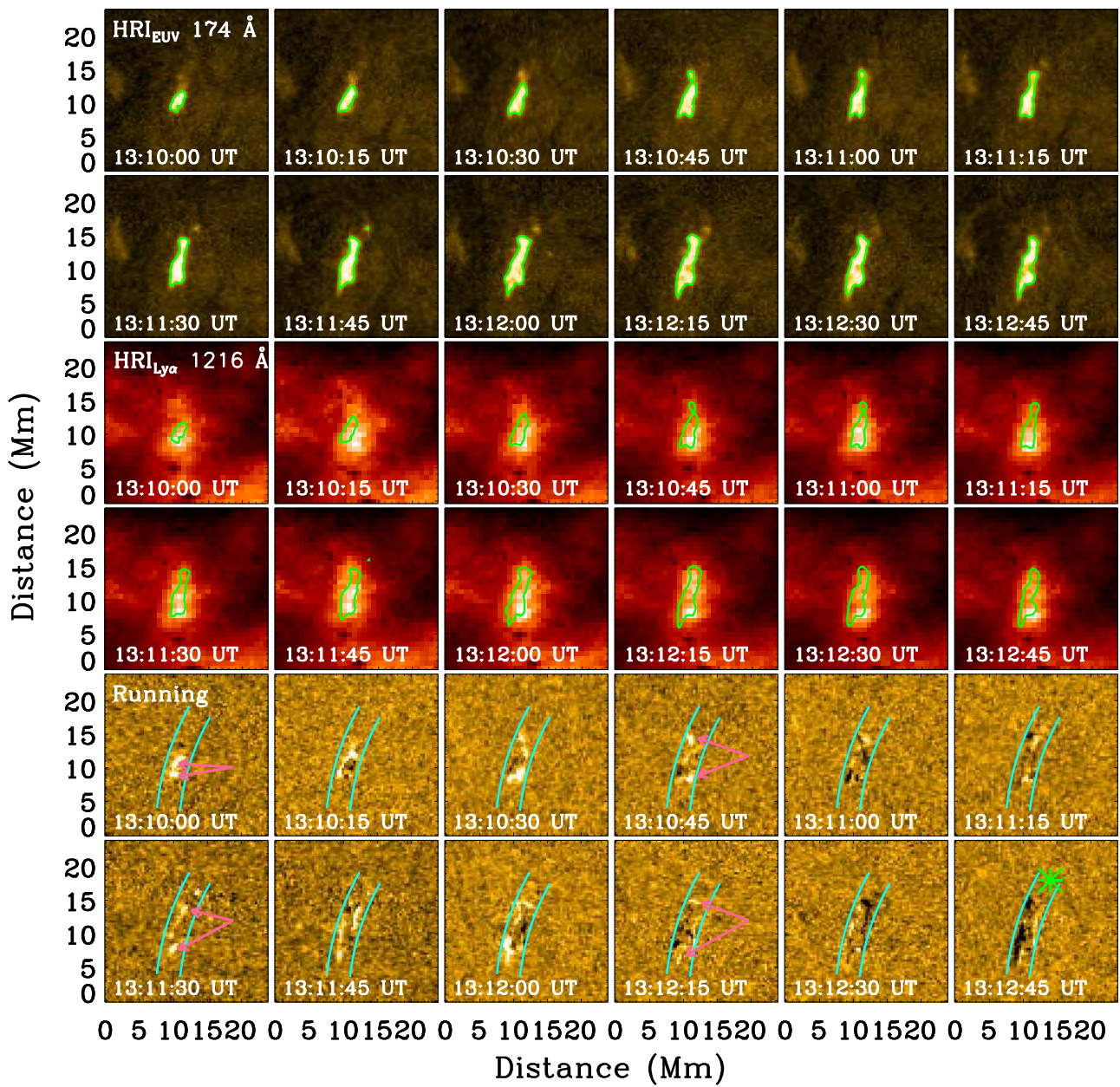}
\caption{Time sequence images with a small FOV of about
24~Mm~$\times$~24~Mm in HRI$_{\rm EUV}$ (top 12 panels) and
HRI$_{\rm Ly\alpha}$ (middle 12 panels) passbands, and running
difference images (bottom 12 panels) in the HRI$_{\rm EUV}$
passband. The green contours represent the HRI$_{\rm EUV}$ emission
at the level of 1000~DN~s$^{-1}$. Two cyan curves mark the curved
slits along the UBP, and the green symbol of `$\ast$' indicate the
zero of y-axis in Figure~\ref{slc1}. The magenta arrows outline
moving structures. \label{bp1}}
\end{figure*}

We analyze the Level~2 (L2) data from the EUI Data
Release~3.0\footnote{https://doi.org/10.24414/k1xz-ae04}, which are
distributed by the EUI team. We perform a cross correlation to
co-align HRI$_{\rm EUV}$ and HRI$_{\rm Ly\alpha}$ images, as shown
in Figure~\ref{snap}. Fifteen UBPs are selected to investigate their
kinetic behaviors, as indicated by numbers. It can be seen that most
of these UBPs are located in the quiet-Sun region, while only three
of them are situated inside (BP8 and BP10) or periphery of (BP3) a
small active region (AR). Panel~(a) shows that two UBPs (BP8 and
BP10) appear almost in the same position but at different times,
suggesting that they are homologous UBPs. It appears that all these
UBPs are located at the boundary of chromospheric network as seen in
the Ly$\alpha$ image, as shown in Figure 1 (b). This is also a
general characteristic of campfires \citep[see][]{Berghmans21}. We
wanted to state that EUI images are easily saturated for the
large-scale phenomenon, i.e., coronal loops. Here, we investigate
the small-scale UBPs, and there are only a few saturated pixels,
which has little impact on our results.

\section{Results}
\subsection{Diverging motions of bi-directional moving structures}
\begin{figure}[ht]
\centering
\includegraphics[width=\linewidth,clip=]{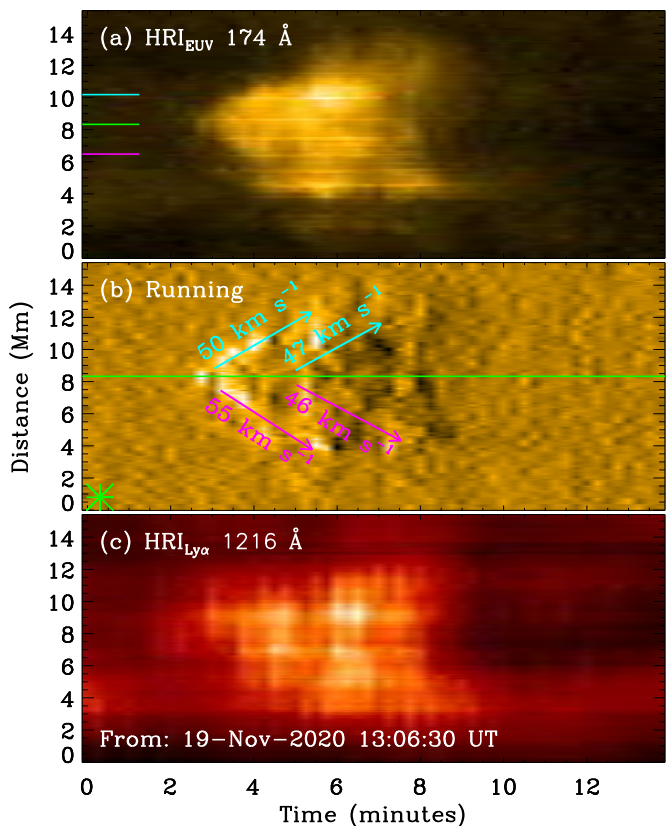}
\caption{Time-distance slices for the BP1 derived from SolO/EUI
image series in the HRI$_{\rm EUV}$ passband (a) and its derivative.
(b), as well as the HRI$_{\rm Ly\alpha}$ passband (c), respectively.
The green line indicates the UBP core, and the cyan and magenta
lines outline the boundaries of the integrated flux in
Figure~\ref{per1}. The cyan and magenta arrows indicate
bi-directional moving structures. \label{slc1}}
\end{figure}

The identified UBPs show loop-like shapes, which are similar to the
profile of CBPs in previous findings
\citep[e.g.,][]{Zhang13,Ning14,Li16}. As an example, we plot time
sequence images of one UBP (labeled `BP1') in the quiet-Sun region
to display its evolution, as shown in Figure~\ref{bp1}. The top 12
panels show intensity images in HRI$_{\rm EUV}$ 174~{\AA} from
13:10:00~UT to 13:12:45~UT. It can be seen that some bright
structures appear in these EUV~174~{\AA} images. They begin with a
bright patch, and then they are almost simultaneously expanding
outward to two ends, for instance, they reveal a diverging motion.
The bright structures expand along not a straight but a curved path,
implying the loop-like shape. The middle 12 panels are intensity
images in HRI$_{\rm Ly\alpha}$ 1216~{\AA} during nearly the same
time interval. There are also some bright structures in these
Ly$\alpha$ 1216~{\AA} images, which are underlying the UBP
identified in EUV~174~{\AA}, as indicated by the green contour.
These bright structures in Ly$\alpha$~1216~{\AA} seem to expand
slowly too. The bottom 12 panels draw running difference images in
HRI$_{\rm EUV}$ 174~{\AA}. In which, we can see that a bright
structure is separated into white and dark kernels, and the white
kernel is always followed by a dark one. Moreover, they are
simultaneously moving toward two ends, as shown by the magenta
arrows, which is consistent with the expanding bright emission in
intensity images. Therefore, we regarded them as moving structures
\citep[cf.][]{Ning14,Li16,Ning20}.

In order to study the moving structures in detail, we plot
time-distance slices along curved slits. The slits are outlined
along a curved shape with two cyan lines in Figure~\ref{bp1}. A
constant width of about 4~Mm is used, so that the bulk of this UBP
brightness can be covered as much as possibly during its whole
lifetime. Next, intensities between two cyan curves are integrated
at every observed time. Then, the time-distance slice is obtained in
HRI$_{\rm EUV}$ 174~{\AA}, as shown in Figure~\ref{slc1}~(a), the
zero of y-axis is indicated by the green symbol of `$\ast$' in
Figure~\ref{bp1}. As can be seen, it becomes bright at roughly
13:09:00~UT, it expands and significantly enhances nearly
simultaneously towards two opposite directions. It gradually
disappears at about 13:16:00~UT. Thus the UBP lasts for
$\sim$7~minutes, and the simultaneous brightenings could be regarded
as bi-directional moving structures from the UBP bright core, as
marked by the green line. The UBP size is estimated by the maximal
length of the bright emission along the y-axis, which is
$\sim$8.5~Mm. Panel~(b) shows the time-distance slice derived from
their running difference images. The moving structures are
identified as oblique streaks, as labeled by the cyan and magenta
arrows. In our observation, the moving structures are separated into
white streaks followed by dark ones, which are consistent with the
diverging motion of the white and dark kernels in running difference
images. Their projected speeds are estimated to be about
46$-$55~km~s$^{-1}$, which are roughly equal. The moving structures
always appear simultaneously, symmetrically and in pairs, as
indicated by the green line and bi-directional arrows. In panel~(c),
we present the time-distance slice in HRI$_{\rm Ly\alpha}$
1216~{\AA}, which shows a weak smooth brightness distribution during
the UBP lifetime. We do not find similar moving structures, most
likely due to the lower spatial resolution of the HRI$_{\rm
Ly\alpha}$ images.

\begin{figure}[ht]
\centering
\includegraphics[width=\linewidth,clip=]{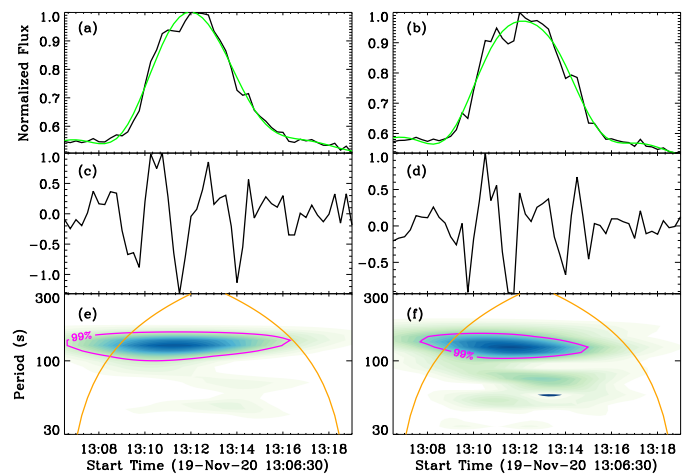}
\caption{Wavelet analysis results for the BP1. Top: Normalized light
curves (black) integrated between two lines (such as between green
and cyan lines for a, between green and magenta lines for b) in
Figure~\ref{slc1}~(a), and slow-varying components (green). Middle:
Rapidly-varying components. Bottom: Morlet wavelet spectra
of the rapidly-varying components in panels (c) and (d), a
significance level of 99\% is outlined by the magenta line.
\label{per1}}
\end{figure}

\begin{figure}[ht]
\centering
\includegraphics[width=0.9\linewidth,clip=]{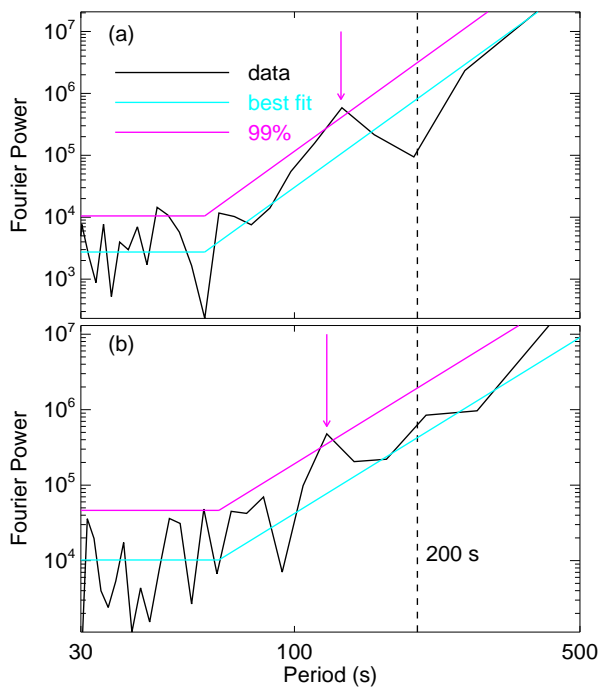}
\caption{Fourier power spectra of the raw light curves. The cyan
line shows the best-fit result, and the magenta line represents the
confidence level at 99\%. The vertical dashed line mark the cutoff
threshold of 200~s. \label{per_fft}}
\end{figure}

The bi-directional moving structures appear to be intermittent and
periodic. Thus, we perform a Morlet wavelet analysis
\citep{Torrence98}, as shown in Figure~\ref{per1}. Panels~(a) \& (b)
plot normalized light curves (black) integrated over two arbitrary
symmetrical locations, as indicated by the green and cyan (or
magenta) lines in Figure~\ref{slc1}~(a). In order to decompose the
slow- (green) and fast- varying components from the raw light curve,
a Fast Fourier Transform (FFT) technique is applied
\citep{Ning17,Li21}. In this case, a cutoff threshold of 200~s is
used, since the lifetime is only 7~minutes. Generally, the
quasi-periodicity refers to at least three successive peaks in the
observed light curve \citep[see,][for
reviews]{Nakariakov19,Zimovets21}, as there is no reason to discuss
the quasi-periodicity when only 1 or 2 emission peaks are detected,
which might be just a coincidence. Therefore, only the quasi-period
less than 200~s is considered. In panels~(c) and (d), we draw the
fast-varying components, and they clearly show three successive
peaks during $\sim$13:09$-$13:15~UT, suggesting a quasi-period of
$\sim$120~s. The quasi-periods are confirmed by the Morlet wavelet
power spectra in panels~(e) and (f). They both show an enhanced
wavelet power during almost the same time. The bulk of two power
spectra are centered at about 120~s, suggesting a dominant period of
$\sim$120~s.

The Morlet wavelet analysis is performed for the fast-varying
component, which strongly depends on the cutoff threshold and might
result into an artificial signal \citep{Kupriyanova10,Auchere16}. To
unambiguously conclude on the quasi-periodicity, we further plot the
Fourier power spectrum of the raw light curve with the Lomb-Scargle
periodogram \citep{Scargle82}, as shown in Figure~\ref{per_fft}.
Similar to previous findings \citep{Vaughan05,Kolotkov18,Li20,Li22},
the Fourier power spectrum is dominated by a power law at the
long-period range and a flat spectrum at the short-period end, as
outlined by the cyan line. A peak at about 120~s is seen to exceed
the 99\% significance level, as indicated by the magenta arrow and
line. Moreover, the 120~s-peak is far from the cutoff threshold, as
marked by the vertical line. Those results demonstrate that the
120-s periodicity is not artificial. The similar quasi-period
confirms that the bi-directional moving structures are symmetrical.
In a word, the moving structures detected in this UBP (BP1) are
characterized by bi-directions, simultaneity, symmetry, and
quasi-periodicity, similarly to what observed with the moving
structures seen in CBPs \citep[cf.,][]{Ning14,Li16}. The diverging
motions of bi-directional moving structures are also observed in
other eight UBPs, as listed in Appendix~\ref{bi_str} and
table~\ref{tab2}.

\subsection{Converging motions between the moving structures}
\begin{figure}[ht]
\centering
\includegraphics[width=\linewidth,clip=]{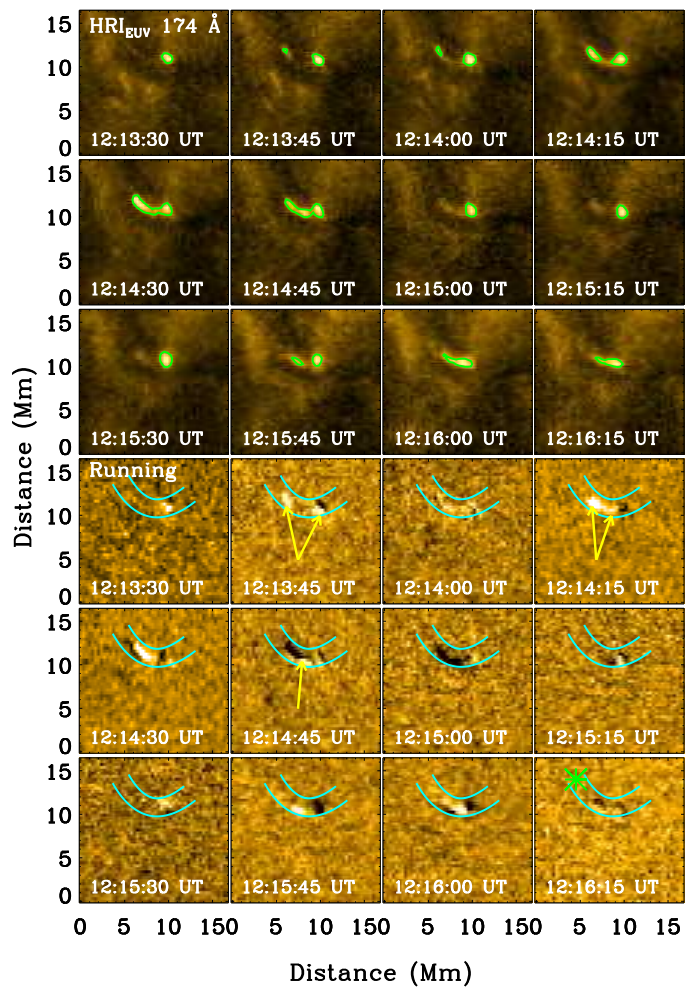}
\caption{Time sequence images with a small FOV of about
17~Mm~$\times$17~Mm in the HRI$_{\rm EUV}$ (top 12 panels) passband,
and their running difference images (bottom 12 panels). The green
contours represent the HRI$_{\rm EUV}$ emission at the level of
900~DN~s$^{-1}$. Two cyan curves mark curved slits along the UBP,
and the green symbol of `$\ast$' indicate the zero of y-axis in
Figure~\ref{slc2}. The yellow arrows indicate moving structures.
\label{bp2}}
\end{figure}

In this case, we investigated a UBP (labeled `BP2') showing
converging motions between a pair of moving structures, as shown in
Figure~\ref{bp2}. Different from the `BP1', it firstly brightens at
about 12:13:45~UT at two separated positions, which might be
regarded as the two footpoints of a loop structure
\citep[cf.,][]{Mandal21}. Then they gradually move to their center
position such as the loop top along a curved path between
12:14:00$-$12:14:45~UT. They finally merge together and exhibit a
loop-like elongated shape, i.e., at about 12:14:45~UT. The moving
structures can be clearly seen in the running difference images,
which are shown by pairs of white and dark kernels, as indicated by
the yellow arrows.

\begin{figure}[ht]
\centering
\includegraphics[width=\linewidth,clip=]{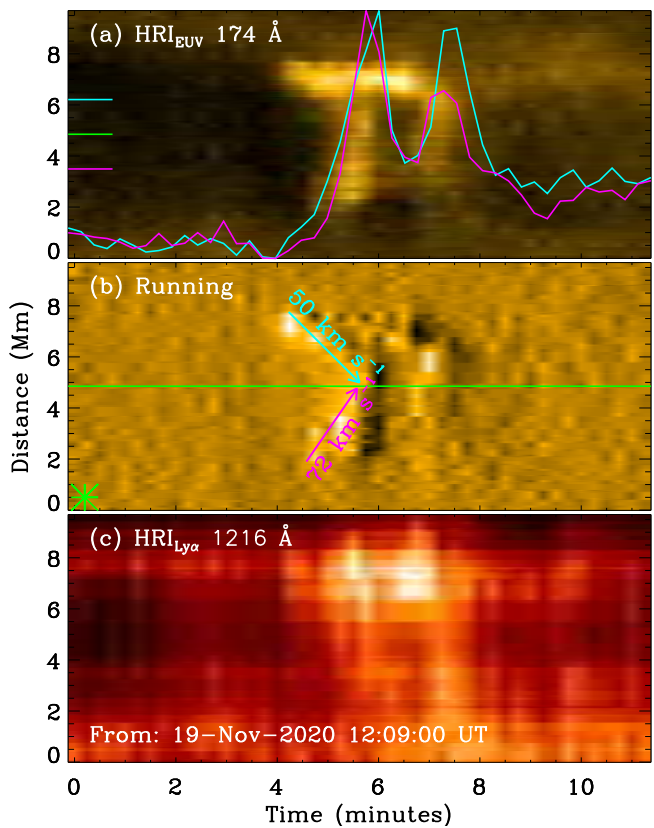}
\caption{Time-distance slices for the BP2 made from SolO/EUI image
series in the HRI$_{\rm EUV}$ passband (a) and their derivative
(gradient along the time axis) (b). The cyan and magenta arrows
indicate the converging motion of two moving structures.
\label{slc2}}
\end{figure}

Figure~\ref{slc2} presents time-distance slices in HRI$_{\rm EUV}$
174~{\AA} (a) and its running difference images (b), as well as in
HRI$_{\rm Ly\alpha}$ 1216~{\AA} (c). This UBP begins to brighten at
two different locations, i.e., two footpoints. The maximum distance
between the two footpoints is taken as the size of this UBP, which
is about 5.5~Mm. Then, the brightenings at the two locations move
closer and then merge into a single source at their center position,
which may be regarded as the converging motion of bi-directional
moving structures. The converging motions repeated twice from about
12:13:15~UT to 12:16:45~UT, and its lifetime is roughly equal to
3.5~minutes. The overplotted light curves in panel~(a) are
integrated over from the short green and cyan/mengta lines in the
left panel. They both show double peaks, and appear almost
simultaneously. However, it is not necessary to discuss the
quasi-periodicity when there are only two peaks. Thus, we did not
perform the wavelet analysis. In panel~(b), moving structures are
identified as oblique streaks, they appear in pairs and show the
converging motion, as indicated by the cyan and magenta arrows.
Their average speeds are estimated to be about 50/72~km~s$^{-1}$.
Panel~(c) reveals a weak brightness distribution in
Ly$\alpha$~1216~{\AA}, suggesting the UBP has a HRI$_{\rm Ly\alpha}$
counterpart. The converging motions of moving structures are also
detected in other four UBPs, which are located in the quiet-Sun
region, see detailed in Appendix~\ref{merg} and table~\ref{tab2}.

\subsection{Sympathetic BP induced by the primary UBP}
\begin{figure}[ht]
\centering
\includegraphics[width=\linewidth,clip=]{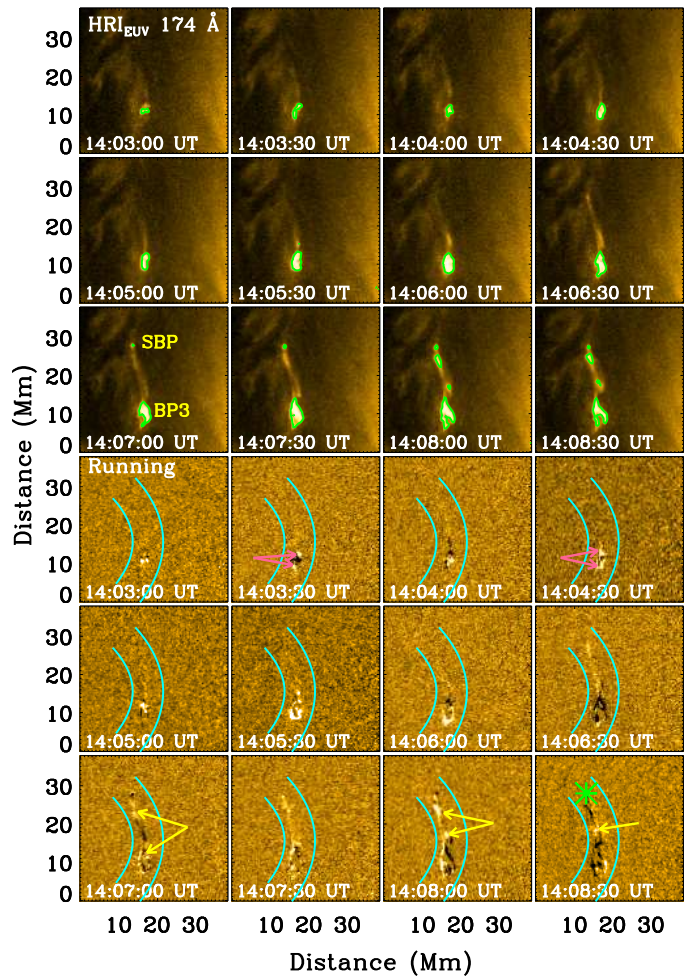}
\caption{Similar to Figure~\ref{bp2}, but with the analysis
performed for the BP3 with a FOV of about 38~Mm~$\times$~38~Mm, The
contour level is set as 1000~DN~s$^{-1}$. \label{bp3}}
\end{figure}

Here, the bi-directional moving structures are observed in the
primary UBP (BP3), and the converging motion can be seen between the
primary UBP and its sympathetic bright point (SBP). Figure~\ref{bp3}
presents time sequence images, showing the temporal evolution of the
UBP (labeled `BP3'). The top 12 panels give intensity images in
HRI$_{\rm EUV}$ 174~{\AA}. Similar to the BP1, the primary UBP
starts with a bright patch, and it expands toward two opposite
directions along a curved path. Then, a remote brightening appear at
about 14:07:00~UT, and it is linked to the primary UBP with a large
loop-like structure, which could be regarded as the SBP
\citep[cf.,][]{Zhang13}. After that, a portion of the primary UBP
and the SBP move closely along the large loop-like path, showing a
converging motion. The moving structures can be clearly seen in
running difference images, as shown in the bottom 12 panels. They
are divided into white and dark kernels, as indicated by the magenta
(diverging motions) and yellow (converging motions) arrows.

\begin{figure}[ht]
\centering
\includegraphics[width=\linewidth,clip=]{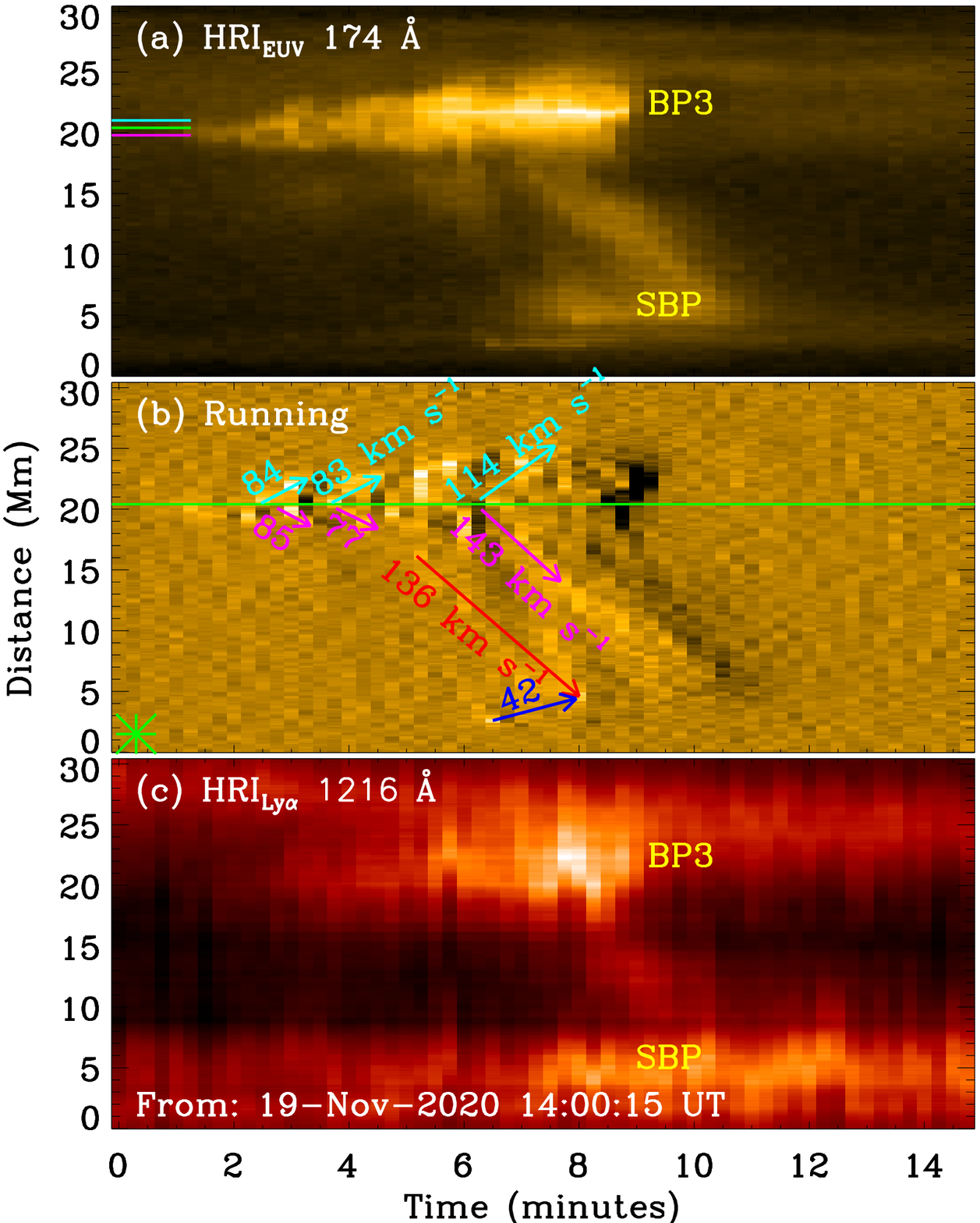}
\caption{Similar to Figure~\ref{slc2}, but with the analysis is
performed for the BP3. \label{slc3}}
\end{figure}

\begin{figure}[ht]
\centering
\includegraphics[width=\linewidth,clip=]{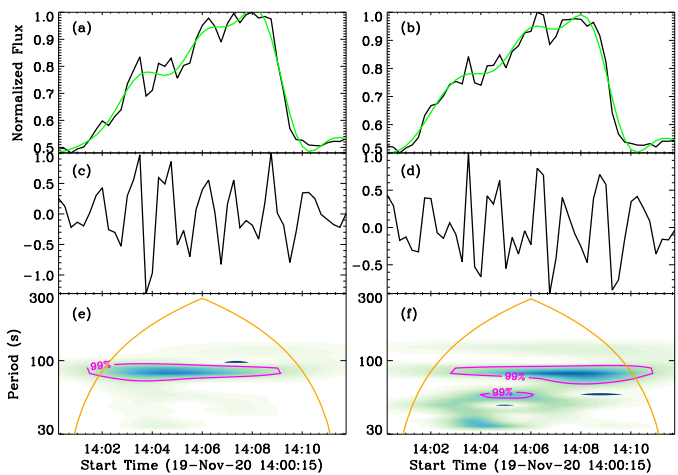}
\caption{Similar to Figure~\ref{per1}, but with the Morlet wavelet
analysis is performed for the BP3. \label{per3}}
\end{figure}

Figure~\ref{slc3} presents time-distance slices derived from the
curved slits, as outlined by two cyan lines in Figure~\ref{bp3}. The
primary UBP can be seen from about 14:01:00~UT to 14:09:00~UT in the
intensity slice (panel~a), which lasts for about 8.0~minutes. The
maximal length of the bright emission is estimated to be about
11.2~Mm. Notice that only the size and lifetime of the primary UBP
is considered here, because the SBP is much weaker than the primary
UBP. The projected speeds are estimated from oblique streaks in the
running difference slice, as shown in panel~(b). At first, the
projected speeds are estimated to be about 77$-$85~km~s$^{-1}$,
which are roughly equal. However, they are a little different when
the SBP appears, such as 114~km~s$^{-1}$ and 143~km~s$^{-1}$.
Similarly, the apparent speeds for converging motions are also
different, which are 136~km~s$^{-1}$ and 42~km~s$^{-1}$. The speed
from the primary UBP (red arrow) is much faster than that from the
SBP (blue arrow). Moreover, the evolution of the primary UBP
and its SBP could be seen in Ly$\alpha$~1216~{\AA}, as shown in
Figure~\ref{slc3}~(c), implying that they have HRI$_{\rm Ly\alpha}$
counterparts.

To look closely the quasi-periodicity, the Morlet wavelet analysis
method is applied to the primary UBP, as shown in Figure~\ref{per3}.
Panels~(a) and (b) plots normalized light curves integrated over two
nearly symmetrical regions. They both show several subpeaks, which
could be regarded as quasi-periodic pulsations. Then, the raw light
curves are decomposed into slow- (green) and fast- (panels c \& d)
varying components by using the FFT technique. Panels~(e) and (f)
present wavelet power spectra of fast-varying components, and they
both show a broad range of periods but centered at about 80~s. This
implies that bi-directional moving structures in the primary UBP are
intermittent and quasi-periodic.

\begin{table}[h]
\addtolength{\tabcolsep}{1pt}
\renewcommand{\arraystretch}{1.0}
\centering
  \caption{Observational parameters of the analyzed UBPs.}
  \label{tab2}
\begin{tabular}{cccccc}
\hline \hline
UBPs   &  Size     &  Duration   & Period & Region &  Motion  \\
       &  (Mm)     &  (minutes)  &  (s)   &        &          \\
\hline
01 &  $\sim$8.5    &  $\sim$7.0  &  120 & quiet  & diverging  \\
02 &  $\sim$5.5    &  $\sim$3.5  &  --  & quiet  & converging \\
03 &  $\sim$11.2   &  $\sim$8.0  &  80  & AR     & diverging  \\
04 &  $\sim$7.3    &  $\sim$8.5  &  130 & quiet  & diverging  \\
05 &  $\sim$6.8    &  $\sim$6.5  &  85  & quiet  & diverging  \\
06 &  $\sim$4.1    &  $\sim$5.0  &  70  & quiet  & diverging  \\
07 &  $\sim$4.6    &  $\sim$4.0  &  --  & quiet  & diverging  \\
08 &  $\sim$31.5   &  $\sim$12.0 &  120 & AR     & diverging  \\
09 &  $\sim$2.8    &  $\sim$3.0  &  --  & quiet  & diverging  \\
10 &  $\sim$26.3   &  $\sim$10   &  75  & AR     & diverging  \\
11 &  $\sim$9.2    &  $\sim$3.0  &  --  & quiet  & diverging  \\
12 &  $\sim$4.6    &  $\sim$4.5  &  --  & quiet  & converging \\
13 &  $\sim$5.1    &  $\sim$6.0  &  --  & quiet  & converging \\
14 &  $\sim$5.7    &  $\sim$4.0  &  --  & quiet  & converging \\
15 &  $\sim$6.8    &  $\sim$3.5  &  --  & quiet  & converging \\
\hline \hline
\end{tabular}
\end{table}

\subsection{Chromospheric responses of UBPs}
\begin{figure}[ht]
\centering
\includegraphics[width=\linewidth,clip=]{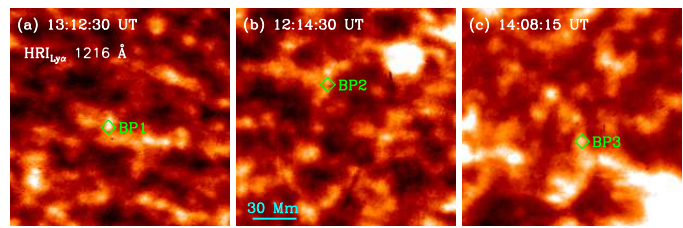}
\caption{Snapshots with a zoomed FOV of about 150~Mm~$\times$~150~Mm
in HRI$_{\rm Ly\alpha}$~1216~{\AA}. The green diamonds mark the
locations of three UBPs, respectively. The cyan tick indicates a
spacing of 30~Mm on the Sun \label{lya}}
\end{figure}

Figure~\ref{lya} shows Ly$\alpha$ images with a zoomed FOV of about
150~Mm~$\times$~150~Mm at three instances of time during the UBPs
labeled as BP1$-$BP3. It can be seen that these three Ly$\alpha$
images exhibit chromospheric network structures, and three UBPs are
located on network boundaries, as outlined by the green diamonds.
That is, when the UBP flash occurs in the HRI$_{\rm EUV}$ passband,
the Ly$\alpha$ emission in the chromosphere is also enhanced, for
instance, showing the Ly$\alpha$ transient brightening. Similarly to
what seen in Ly$\alpha$ images of these three UBPs, other twelve
UBPs are found to have Ly$\alpha$ brightenings, and they appear to
locate at the location of network boundaries, as shown in
Figures~\ref{a_lya} and \ref{b_lya}.

\section{Conclusion and Discussion}
Based on high-spatial resolution data sets measured by SolO/EUI in
passbands of HRI$_{\rm EUV}$ and HRI$_{\rm Ly\alpha}$, we explore
moving structures in fifteen UBPs both in the quiet-Sun region and
in the AR. Two classes of moving structures are found: (I) Diverging
motions of bi-directional moving structures are observed in ten
UBPs; (II) Converging motions between double moving structures are
seen in five UBPs. Besides, we found a SBP that is induced by the
primary UBP, and the converging motion along a large loop-like
structure is found between their moving structures. The moving
structures observed in UBPs seem to be similar to the propagating
brightenings reported by \cite{Mandal21}. However, the merger-type
event of propagating brightenings started at one footpoint and moved
toward another footpoint, and their repeated plasma ejections showed
the uni-directional motion \cite[see][]{Mandal21}. Our UBPs exhibit
bi-directional movements originated from a bright core and propagate
toward two ends, the converging motions begin simultaneously at two
footpoints and moved toward the loop-top region. We do not find the
reflected motion. This is mainly because that the EUI data used by
\cite{Mandal21} has nearly a factor of two higher
spatial-resolutions and three higher cadences.

Using higher-cadence observations from SolO/EUI, several groups of
small-scale UV brightenings have been reported in the quiet-Sun
region, i.e., campfires \citep{Berghmans21}, microjets
\citep{Hou21}, and fast repeating jets \citep{Chitta21}, et cetera.
Campfires appear as transient UV brightenings, whereas they do not
reveal apparently localized Ly$\alpha$ brightenings. Their length
sizes are smaller than 4~Mm and their lifetimes are shorter than
200~s \citep{Berghmans21,Chen21}. Microjets are identified as
collimated structures with bright emissions at footpoints and
extended outward spires, some of them show pronounced signatures in
the HRI$_{\rm Ly\alpha}$ passband. They have the maximum length
scale of 7.7~Mm, an average duration of 4.6~minutes, and a projected
speed of 62~km~s$^{-1}$ \citep{Hou21}. Fast repeating jets show
intermittent jet activities on a very short timescale with
uni-direction or bi-directions, while their propagating speeds can
be as fast as about 150~km~s$^{-1}$ \citep{Chitta21}. All those UV
brightenings are transient in nature, and they are proposed to be
generated by the small-scale magnetic reconnection and could
contribute to the coronal heating, particularly in the quiet Sun
\citep[e.g.,][]{Chen21,Panesar21,Zhukov21}. In this study, UBPs seem
to have larger sizes for longer lifetimes, and more than 1/3 of them
have duration below 240~s (Table~1). Campfires found by
\cite{Berghmans21} are limited by their short duration, and there
are certainly many events with duration longer than 200~s. Thus,
UBPs and campfires most likely share the similar magnetic origin,
e.g., small-scale reconnection at coronal heights
\citep{Chen21,Panesar21}. Our observations extend the distribution
of small UV bright events, such as, UBPs, campfires, etc. The length
size and lifetime are roughly equal to that observed in microjets,
but the microjet often shows the collimated outflow along an open
magnetic field line \citep{Hou21}. The bi-directional moving
structures of UBPs are mostly confined in the closed magnetic loops,
which are similar to fast repeating jets reported by
\cite{Chitta21}. However, those fast repeating jets have very short
lifetimes and quick speeds, mainly due to the much higher temporal
resolution of EUI data. Our UBPs reveal bi-directional moving
structures, whether they are divergence or convergence. They all
show obviously signatures in Ly$\alpha$~1216~{\AA} images and appear
to locate on network boundaries, suggesting that they have responses
both in the corona and chromosphere.

Bi-directional moving structures originated from the bright core
have been investigated in two CBPs and a C-class flare at multiple
wavelengths \citep{Ning14,Ning16,Li16}. They are characterized by
bi-directions, simultaneity, symmetry, and quasi-periodicity. Their
apparent speeds are estimated to be about 200$-$300~km~s$^{-1}$.
Both the two CBPs and the C-class flare were seen in big ARs. In our
observation, three UBPs are detected inside (BP8 and BP10) or
closely to (BP3) a small AR. Thus, their length sizes, lifetimes and
average apparent speeds are larger than those seven UBPs in the
quiet-Sun region, as shown in table~\ref{tab2}. However, the fastest
speed is only about 140~km~s$^{-1}$, which is still smaller than
previous findings in AR CBPs \cite[e.g.,][]{Ning14,Li16}. This is
might because that the AR in our case is small. The apparent speeds
of seven UBPs in the quiet-Sun region are estimated to
$\sim$33$-$85~km~s$^{-1}$, which are roughly equal to the projected
speeds observed in microjets \citep{Hou21} and propagating
brightenings \citep{Mandal21}, implying their common origin.
Previous observations suggest that these bi-directional moving
structures could be regarded as outflows after the magnetic
reconnection \citep{Ning14,Li16,Tiwari19,Ning20}, this is consistent
with bi-directional plasma flows on a short timescale in fast
repeating jets \citep{Chitta21}. We notice that three UBPs (BP2,
BP7, and BP9) do not reveal quasi-periodicity, largely due to their
short lifetime and lower cadence in our data. We will check the
periodicity with high-cadence HRI$_{\rm EUV}$ images in future
\citep[e.g.,][]{Chitta21,Mandal21}.

In this study, we also find bi-directional moving structures from
two-end sources to their center region along loop-like paths in five
UBPs, i.e., converging motions between two moving structures. The
classical 2-D reconnection model suggests that both bi-directional
inflows and outflows are generated by the magnetic reconnection
\citep{Sturrock64,Priest94,Li19,Peter19}. Therefore, the converging
motions could be considered as bi-directional inflows. Moreover, a
SBP induced by the primary UBP (BP3) is seen at the periphery of a
small AR, and they are moving closely along a large-scale loop path,
merging together in the loop-top source. Such converging motion
between a primary CBP and its SBP is regarded as the signature of
chromosphere evaporation \citep[see,][]{Zhang13}. Therefore, the
converging motions between two moving structures might also be
explained as upflows driven by the chromosphere evaporation.
However, we cannot conclude the nature of their mechanism, because
the lack of other observations, such as magnetic fields, HXR
sources, and spectral lines. And it is impossible to determine the
location of the footpoints and loop top, which is based solely on
the shape and movement of UBPs \citep[see aslo,][]{Mandal21}. For a
more reliable conclusion, joint observations with other remote
sensing instruments with high resolutions are required, for
instance, the Polarimetric and Helioseismic Imager
\citep{Solanki20}, the Spectrometer/Telescope for Imaging X-rays
\citep{Krucker20}, and the Spectral Imaging of the Coronal
Environment \citep{Spice20} on board SolO.

\begin{acknowledgements}
We thank the referee for inspiring comments to improve the quality
of this article. The author would like to appreciate Drs.
D.~Berghmans and F.~Auch\`{e}re for their kindly discussing about
the SolO/EUI data. This study is supported by NSFC under grant
11973092, 12073081, U1931138, 11790302, the Strategic Priority
Research Program on Space Science, CAS, Grant No. XDA15052200 and
XDA15320301. D. Li is also supported by the Surface Project of
Jiangsu Province (BK20211402). Solar Orbiter is a space mission of
international collaboration between ESA and NASA, operated by ESA.
The EUI instrument was built by CSL, IAS, MPS, MSSL/UCL, PMOD/WRC,
ROB, LCF/IO with funding from the Belgian Federal Science Policy
Office (BELSPO/PRODEX PEA 4000112292); the Centre National d'Etudes
Spatiales (CNES); the UK Space Agency (UKSA); the Bundesministerium
f\"{u}r Wirtschaft und Energie (BMWi) through the Deutsches Zentrum
f\"{u}r Luft-und Raumfahrt (DLR) and the Swiss Space Office (SSO).
\end{acknowledgements}

\begin{appendix}
\section{Diverging motions of Bi-directional moving structures}
\begin{figure}[ht]
\centering
\includegraphics[width=\linewidth,clip=]{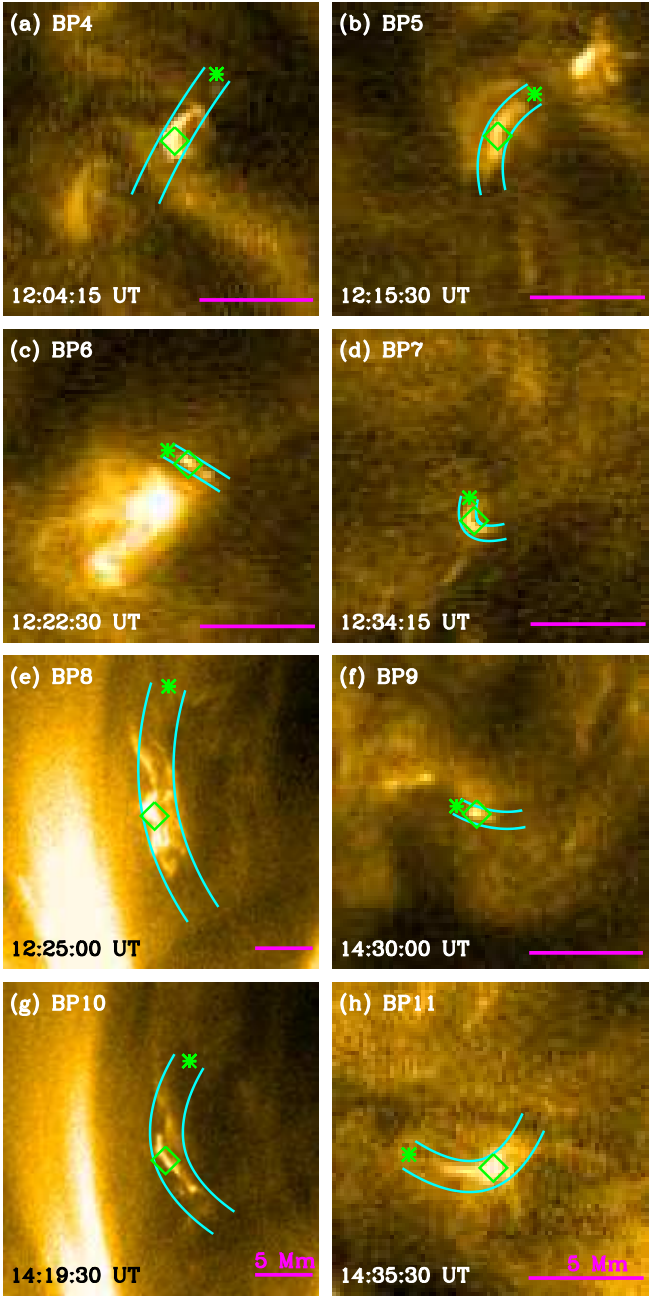}
\caption{Snapshots with a small FOV in the HRI$_{\rm EUV}$ passband.
The FOV is about 14~Mm~$\times$~14~Mm for the BP4$-$BP7, BP9, and
BP11, while the FOV is about 28~Mm~$\times$~28~Mm for BP8 and BP10.
Two cyan lines outline the curved slits, and the green `$\ast$'
indicates the start point. The green diamonds mark UBP locations.
The magenta tick indicates a space of 5~Mm on the Sun.
\label{a_euv}}
\end{figure}

\begin{figure}[ht]
\centering
\includegraphics[width=\linewidth,clip=]{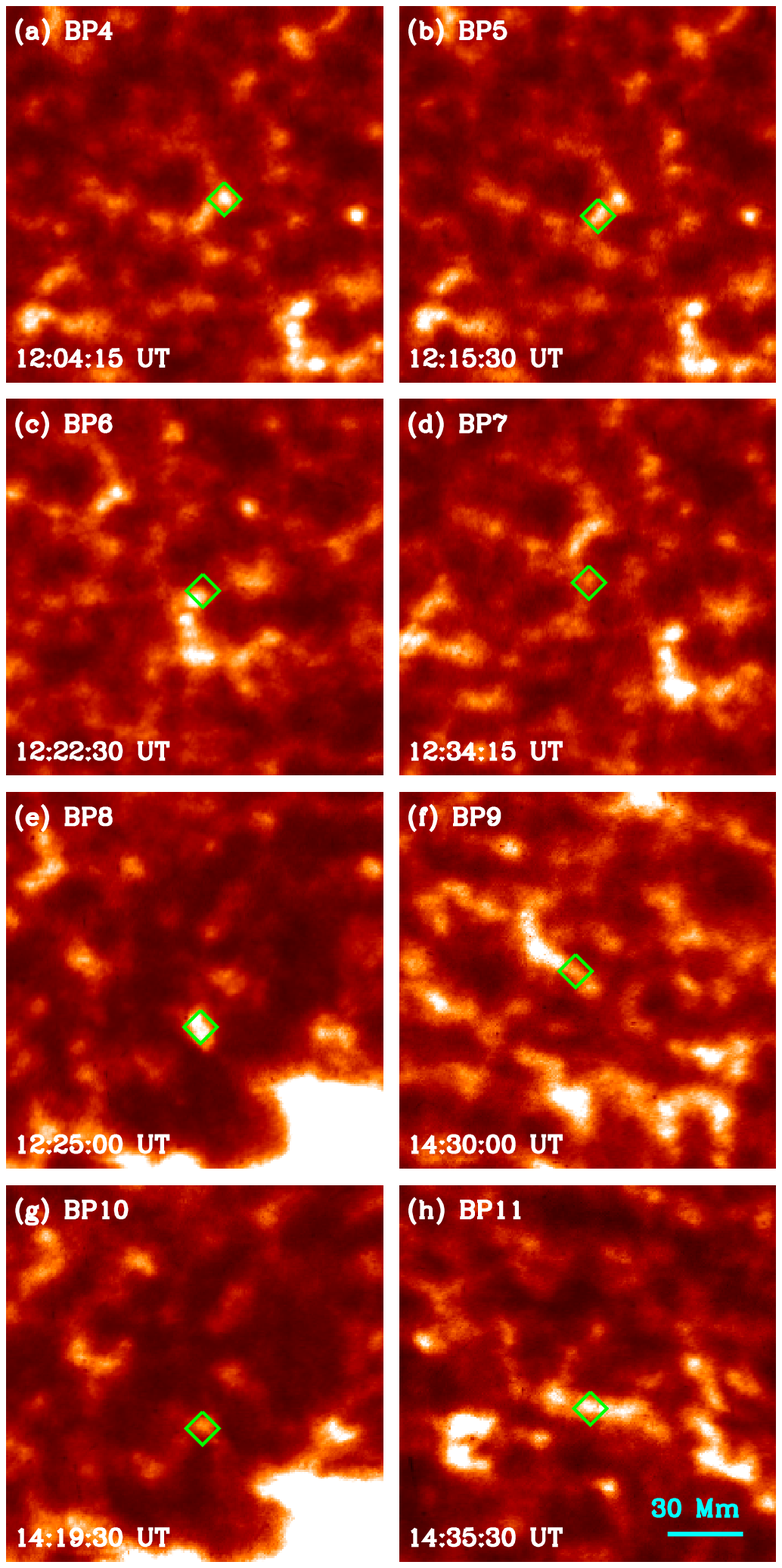}
\caption{Snapshots with a zoomed FOV of about 150~Mm~$\times$~150~Mm
in the HRI$_{\rm Ly\alpha}$ passband for the BP4$-$BP11. The green
diamonds mark UBP locations as shown in Figure~\ref{a_euv}. The cyan
tick indicates a space of 30~Mm on the Sun \label{a_lya}}
\end{figure}

\begin{figure}[ht]
\centering
\includegraphics[width=\linewidth,clip=]{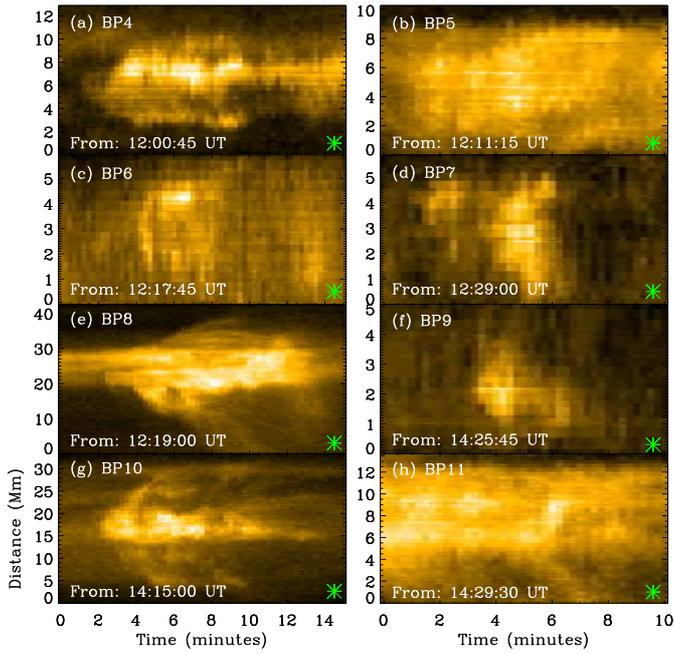}
\caption{Time-distance slices for the BP4$-$BP11 in the HRI$_{\rm
EUV}$ passband. \label{a_slc1}}
\end{figure}

\begin{figure}[ht]
\centering
\includegraphics[width=\linewidth,clip=]{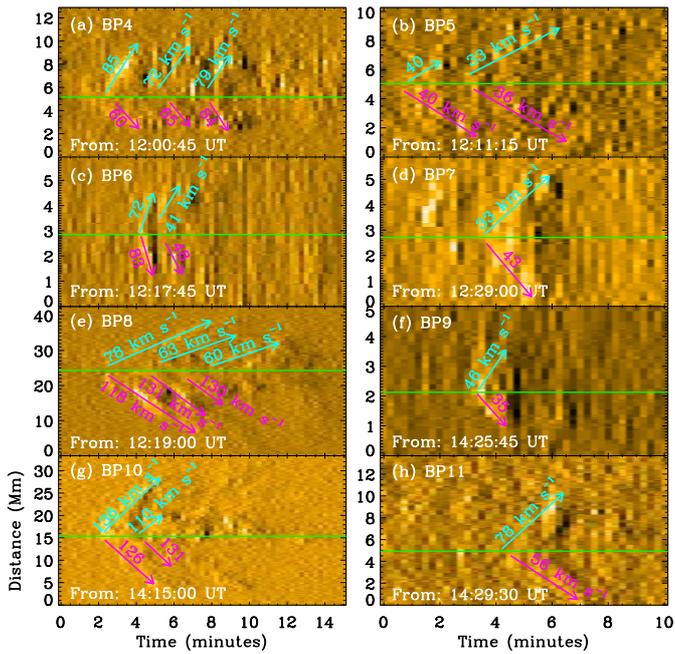}
\caption{Derivative of time-distance slices in Figure~\ref{a_slc1}.
The color arrows indicate moving structures. \label{a_slc2}}
\end{figure}

\label{bi_str} In this section, we show eight UBPs that clearly
exhibit bi-directional moving structures originated from a bright
core and propagated toward two ends. Figures~\ref{a_euv} and
\ref{a_lya} plot snapshots of the BP4$-$BP11 in passbands of
HRI$_{\rm EUV}$ and HRI$_{\rm Ly\alpha}$, respectively. Two cyan
curves outline the slit positions, while the green diamonds mark UBP
locations. Figures~\ref{a_slc1} and \ref{a_slc2} draw time-distance
slices for these UBPs derived from the EUV~174~{\AA} intensity and
their running difference images, respectively. The projected speeds
are also labeled.

\section{Converging motions between two moving structures}
\label{merg} 
\begin{figure}[ht]
\centering
\includegraphics[width=\linewidth,clip=]{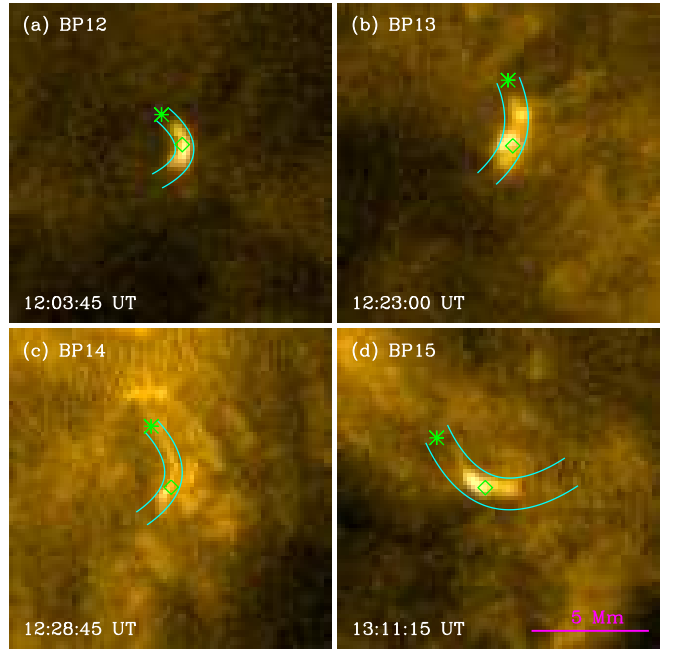}
\caption{Snapshots with a same FOV of about 14~Mm~$\times$~14~Mm for
the BP12$-$BP15 in the HRI$_{\rm EUV}$ passband. Two cyan curves
outline slit positions, and the green `$\ast$' indicates the start
point. The magenta tick indicates a space of 5~Mm on the Sun.
\label{b_euv}}
\end{figure}

\begin{figure}[ht]
\centering
\includegraphics[width=\linewidth,clip=]{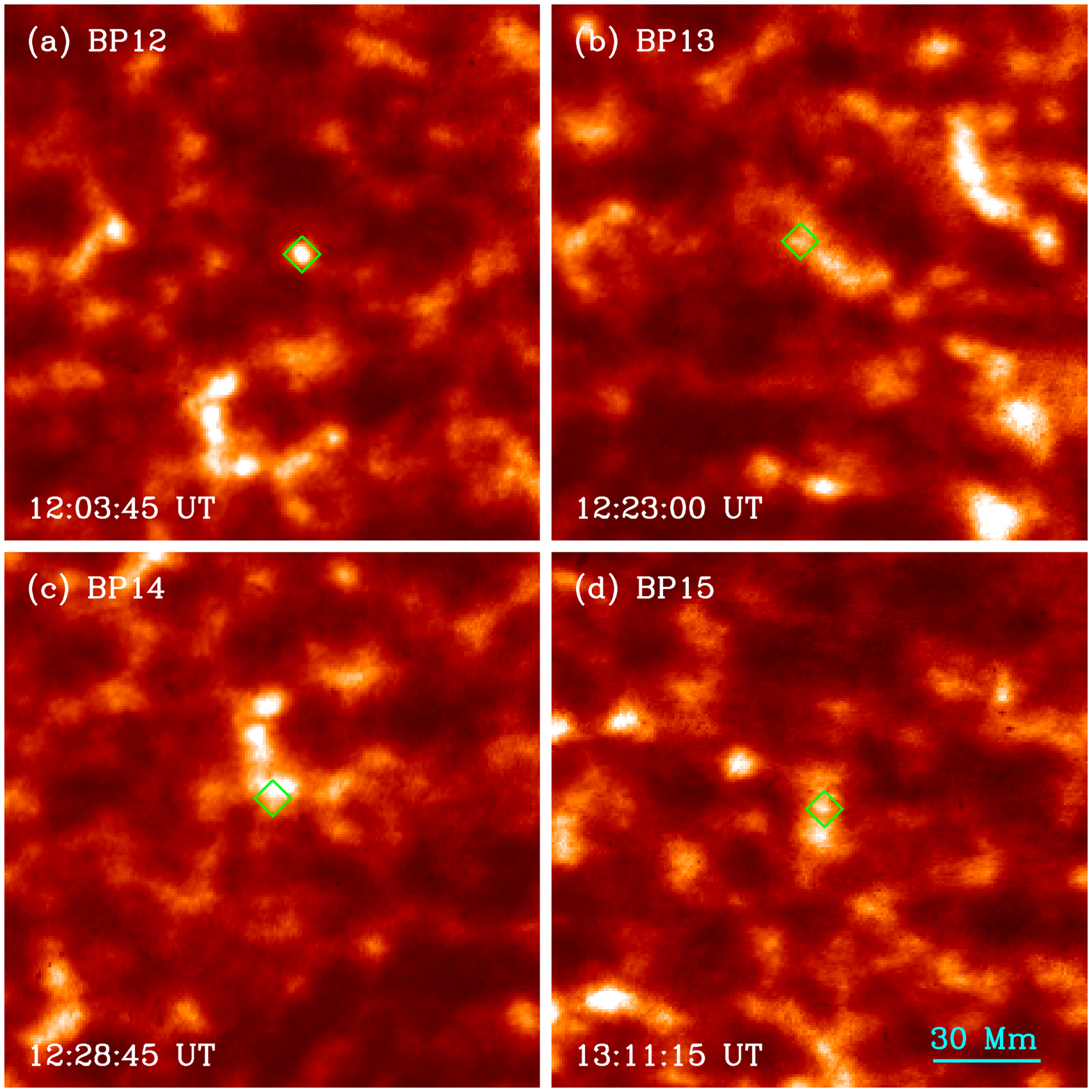}
\caption{Snapshots with a zoomed FOV of about 150~Mm~$\times$~150~Mm
in the HRI$_{\rm Ly\alpha}$ passband for the BP12$-$BP15. The green
diamonds mark UBP locations as shown in Figure~\ref{b_euv}. The cyan
tick indicates a space of 30~Mm on the Sun \label{b_lya}}
\end{figure}

\begin{figure}[ht]
\centering
\includegraphics[width=\linewidth,clip=]{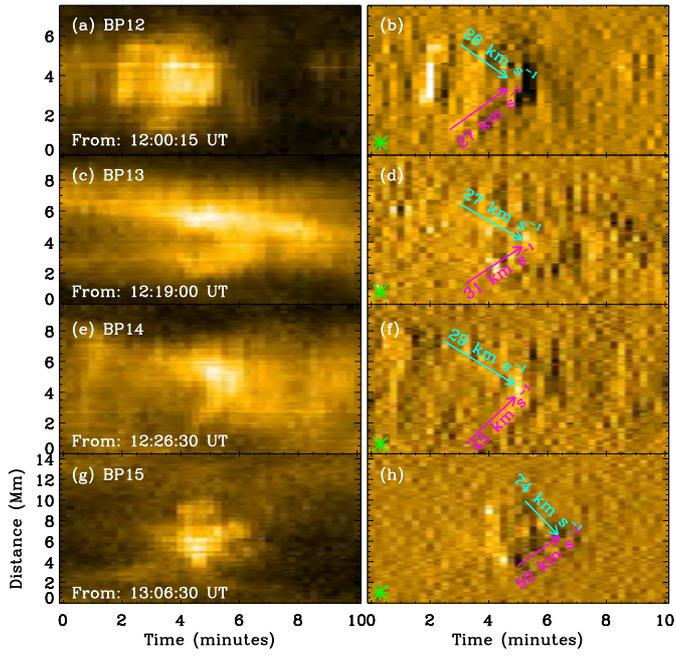}
\caption{Time-distance slices for the BP12$-$BP15 in the HRI$_{\rm
EUV}$ passband (left) and their gradients along the time axis
(right). The color arrows indicate converging motions.
\label{b_slc}}
\end{figure}

In this section, we show four UBPs that clearly exhibit
converging motions from two-end locations to their center regions.
Figures~\ref{b_euv} and \ref{b_lya} plot snapshots of the
BP12$-$BP15 in passbands of HRI$_{\rm EUV}$ and HRI$_{\rm
Ly\alpha}$, respectively. Two cyan curves outline slit positions,
and the green diamonds mark UBP locations. Figure~\ref{b_slc} plots
time-distance slices for these UBPs derived from the
EUV~174~intensity and their running difference images, respectively.
The apparent speeds are also shown.

\end{appendix}
\end{document}